\documentclass[final]{aipproc}

\layoutstyle{8x11double}

\usepackage[utf8x]{inputenc}              % input encoding
\usepackage[T1]{fontenc}                  % font encoding
\usepackage{xcolor}                       % colors
\usepackage{amsfonts}
\usepackage{amsmath}
\usepackage{amssymb}
\usepackage[english]{babel}
\usepackage{graphicx}
\usepackage{units}
\usepackage{url}                          % display of URLS
\usepackage[hang,raggedright]{subfigure}  % subfigures
\usepackage{nicefrac}
\usepackage{textcomp}                     % °Celsius
\usepackage[pdfpagelabels]{hyperref}

\newcommand{\cawo}{CaWO$_{4}$ }
\newcommand{\znwo}{ZnWO$_{4}$ }

\DeclareMathAlphabet{\mathbi}{OML}{cmm}{b}{it}

\allowdisplaybreaks[1]

\definecolor{grey}{rgb}{0.75,0.75,0.75}
\definecolor{brown}{rgb}{0.5,0.25,0.0}
\definecolor{pink}{rgb}{1.0,0.5,0.5}
\definecolor{darkgreen}{rgb}{0,0.5,0}
\definecolor{orange}{rgb}{1,0.5,0}
\definecolor{plotorange}{rgb}{1,0.65,0}
\definecolor{lightgray}{rgb}{0.8,0.8,0.8}
\definecolor{gold}{rgb}{1,0.8,0}
\definecolor{superfluid}{rgb}{0.5,0.75,0.75}
\definecolor{normalliquid}{rgb}{0,0.5,0.5}

\hypersetup{
pdftitle={Composite CaWO4-Detectors for the CRESST-II Experiment},
pdfauthor={Michael Kiefer},
pdfkeywords={Dark Matter, WIMP, Solid-state detectors, Low temperature detectors, Scintillation detectors, CaWO4, Epoxy, Glue}
}

\begin{document}
\title{Composite \cawo Detectors for the CRESST-II Experiment}
\classification{95.35.+d, 29.40.Wk, 07.20.Mc, 29.40.Mc}
\keywords{Dark Matter, WIMP, Low temperature detectors, Scintillation detectors, CaWO${_4}$, Epoxy, Glue}
\author{M.~Kiefer}{address={Max-Planck-Institut f\"ur Physik, F\"ohringer Ring 6, D-80805 M\"unchen, Germany}}
\author{G.~Angloher}{address={Max-Planck-Institut f\"ur Physik, F\"ohringer Ring 6, D-80805 M\"unchen, Germany}}
\author{M.~Bauer}{address={Eberhard-Karls-Universit\"at T\"ubingen, D-72076 T\"ubingen, Germany}}
\author{I.~Bavykina}{address={Max-Planck-Institut f\"ur Physik, F\"ohringer Ring 6, D-80805 M\"unchen, Germany}}
\author{A.~Bento}{address={Max-Planck-Institut f\"ur Physik, F\"ohringer Ring 6, D-80805 M\"unchen, Germany},altaddress={on leave from: Departamento de Fisica, Universidade de Coimbra, P3004 516 Coimbra, Portugal}}
\author{A.~Brown}{address={Department of Physics, University of Oxford, Oxford OX1 3RH, United Kingdom}}
\author{C.~Bucci}{address={INFN, Laboratori Nazionali del Gran Sasso, I-67010 Assergi, Italy}}
\author{C.~Ciemniak}{address={Physik-Department E15, Technische Universit\"at M\"unchen, D-85747 Garching, Germany}}
\author{C.~Coppi}{address={Physik-Department E15, Technische Universit\"at M\"unchen, D-85747 Garching, Germany}}
\author{G.~Deuter}{address={Eberhard-Karls-Universit\"at T\"ubingen, D-72076 T\"ubingen, Germany}}
\author{F.~von~Feilitzsch}{address={Physik-Department E15, Technische Universit\"at M\"unchen, D-85747 Garching, Germany}}
\author{D.~Hauff}{address={Max-Planck-Institut f\"ur Physik, F\"ohringer Ring 6, D-80805 M\"unchen, Germany}}
\author{S.~Henry}{address={Department of Physics, University of Oxford, Oxford OX1 3RH, United Kingdom}}
\author{P.~Huff}{address={Max-Planck-Institut f\"ur Physik, F\"ohringer Ring 6, D-80805 M\"unchen, Germany}}
\author{J.~Imber}{address={Department of Physics, University of Oxford, Oxford OX1 3RH, United Kingdom}}
\author{S.~Ingleby}{address={Department of Physics, University of Oxford, Oxford OX1 3RH, United Kingdom}}
\author{C.~Isaila}{address={Physik-Department E15, Technische Universit\"at M\"unchen, D-85747 Garching, Germany}}
\author{J.~Jochum}{address={Eberhard-Karls-Universit\"at T\"ubingen, D-72076 T\"ubingen, Germany}}
\author{M.~Kimmerle}{address={Eberhard-Karls-Universit\"at T\"ubingen, D-72076 T\"ubingen, Germany}}
\author{H.~Kraus}{address={Department of Physics, University of Oxford, Oxford OX1 3RH, United Kingdom}}
\author{J.-C.~Lanfranchi}{address={Physik-Department E15, Technische Universit\"at M\"unchen, D-85747 Garching, Germany}}
\author{R.~F.~Lang}{address={Max-Planck-Institut f\"ur Physik, F\"ohringer Ring 6, D-80805 M\"unchen, Germany}}
\author{M.~Malek}{address={Department of Physics, University of Oxford, Oxford OX1 3RH, United Kingdom}}
\author{R.~McGowan}{address={Department of Physics, University of Oxford, Oxford OX1 3RH, United Kingdom}}
\author{V.~B.~Mikhailik}{address={Department of Physics, University of Oxford, Oxford OX1 3RH, United Kingdom}}
\author{E.~Pantic}{address={Max-Planck-Institut f\"ur Physik, F\"ohringer Ring 6, D-80805 M\"unchen, Germany}}
\author{F.~Petricca}{address={Max-Planck-Institut f\"ur Physik, F\"ohringer Ring 6, D-80805 M\"unchen, Germany}}
\author{S.~Pfister}{address={Physik-Department E15, Technische Universit\"at M\"unchen, D-85747 Garching, Germany}}
\author{W.~Potzel}{address={Physik-Department E15, Technische Universit\"at M\"unchen, D-85747 Garching, Germany}}
\author{F.~Pr\"obst}{address={Max-Planck-Institut f\"ur Physik, F\"ohringer Ring 6, D-80805 M\"unchen, Germany}}
%\author{W.~Rau}{address={Physik-Department E15, Technische Universit\"at M\"unchen, D-85747 Garching, Germany}, altaddress={now at: Department of Physics, Queen's University, Ontario K7L 3N6, Canada}}
\author{S.~Roth}{address={Physik-Department E15, Technische Universit\"at M\"unchen, D-85747 Garching, Germany}}
\author{K.~Rottler}{address={Eberhard-Karls-Universit\"at T\"ubingen, D-72076 T\"ubingen, Germany}}
\author{C.~Sailer}{address={Eberhard-Karls-Universit\"at T\"ubingen, D-72076 T\"ubingen, Germany}}
\author{K.~Sch\"affner}{address={Max-Planck-Institut f\"ur Physik, F\"ohringer Ring 6, D-80805 M\"unchen, Germany}}
\author{J.~Schmaler}{address={Max-Planck-Institut f\"ur Physik, F\"ohringer Ring 6, D-80805 M\"unchen, Germany}}
\author{S.~Scholl}{address={Eberhard-Karls-Universit\"at T\"ubingen, D-72076 T\"ubingen, Germany}}
\author{W.~Seidel}{address={Max-Planck-Institut f\"ur Physik, F\"ohringer Ring 6, D-80805 M\"unchen, Germany}}
\author{L.~Stodolsky}{address={Max-Planck-Institut f\"ur Physik, F\"ohringer Ring 6, D-80805 M\"unchen, Germany}}
\author{A.~J.~B.~Tolhurst}{address={Department of Physics, University of Oxford, Oxford OX1 3RH, United Kingdom}}
\author{I.~Usherov}{address={Eberhard-Karls-Universit\"at T\"ubingen, D-72076 T\"ubingen, Germany}}
\author{W.~Westphal}{address={Physik-Department E15, Technische Universit\"at M\"unchen, D-85747 Garching, Germany}, altaddress={Deceased}}

\begin{abstract}
CRESST-II, standing for Cryogenic Rare Events Search with Superconducting Thermometers phase II, is an experiment searching for Dark Matter. In the LNGS facility in Gran Sasso, Italy, a cryogenic detector setup is operated in order to detect WIMPs by elastic scattering off nuclei, generating phononic lattice excitations and scintillation light. The thermometers used in the experiment consist of a tungsten thin-film structure evaporated onto the \cawo absorber crystal. The process of evaporation causes a decrease in the scintillation light output. This, together with the need of a big-scale detector production for the upcoming EURECA experiment lead to investigations for producing thermometers on smaller crystals which are glued onto the absorber crystal. In our Run 31 we tested composite detectors for the first time in the Gran Sasso setup. They seem to produce higher light yields as hoped and could provide an additional time based discrimination mechanism for low light yield clamp events.
\end{abstract}
\maketitle
\section{Introduction}
The CRESST-II experiment aims for direct detection of WIMPs via elastic scattering off nuclei of scintillating crystals. The scattering energy of impinging particles is transformed into phonons and scintillation light. The phonons enable us to precisely measure the energy of the particles, while the amount of emitted light enables us to determine their nature \cite{Meunier}.\par
\begin{figure}
  \includegraphics[width=0.9\linewidth]{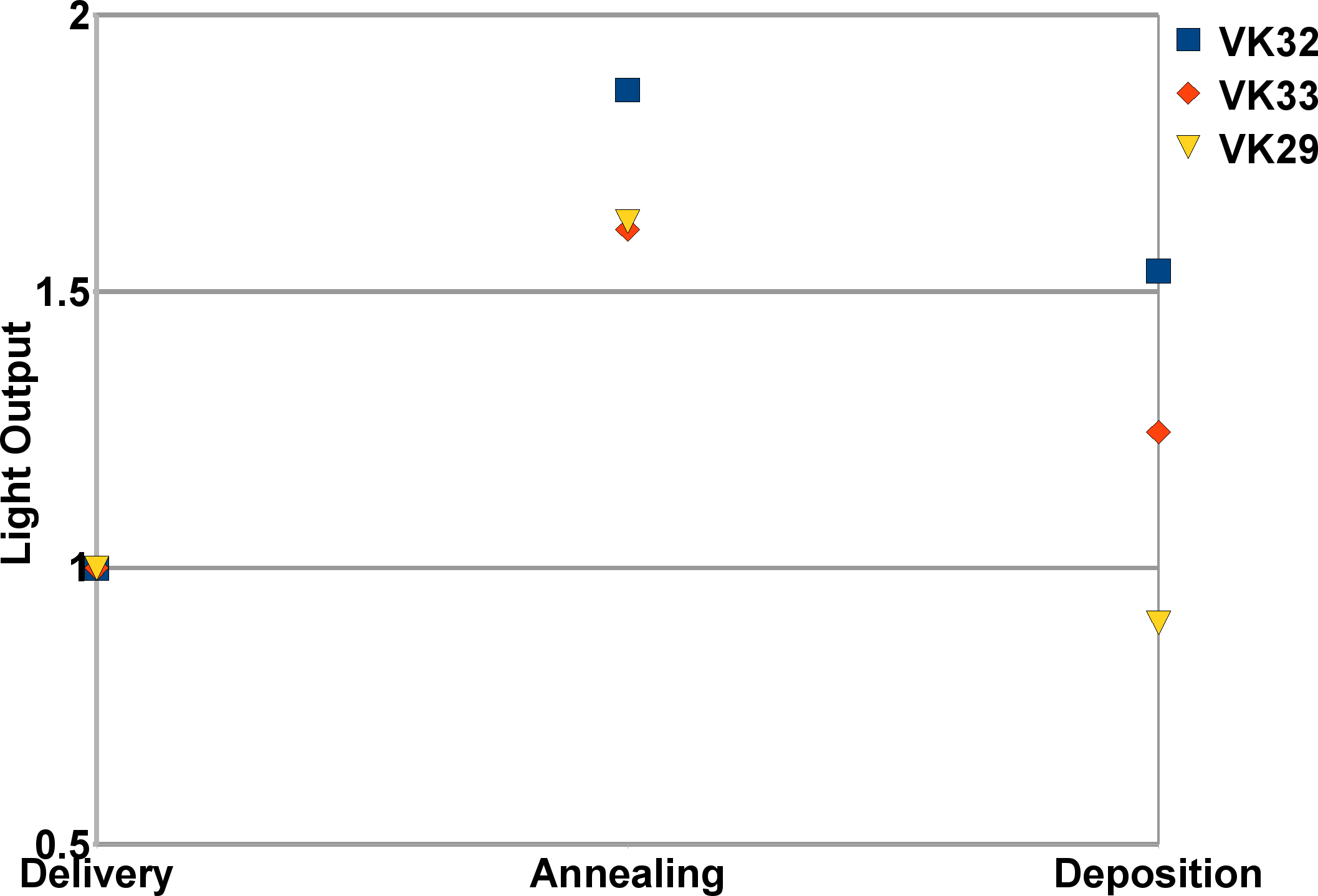}
 \caption{Development of light output of three different crystals during detector production. The values are normalized to the light output at delivery time.}
 \label{fig:lightoutput}
\end{figure}

A CRESST-II detector module consists of a phonon detector and a separate light detector \cite{Angloher2009_run30}. The phonon detector is a cylindrical crystal of scintillating material, while the light detector is a thin wafer optimized for the detection of the scintillation light. The housing around them is covered with scintillating reflector foil. For signal read-out, both crystal and wafer have an evaporated thin film  superconducting phase transition thermometer. The high temperature needed for growing good tungsten films results in a degradation of the light output of the scintillator crystal (see Fig.\ref{fig:lightoutput}).\par
In our previous run \cite{Angloher2009_run30}, we observed 3 events in the WIMP signal acceptance region. An increase of the detected light signal would answer this question: It should help to clarify whether these events are just neutrons misidentified due to a low light output. Our approach was to increase the light output by avoiding the high temperature treatment of the crystals in vacuum and by this mean to prevent the crystals from degrading.\par
\begin{figure}
  \includegraphics[width=0.9\linewidth]{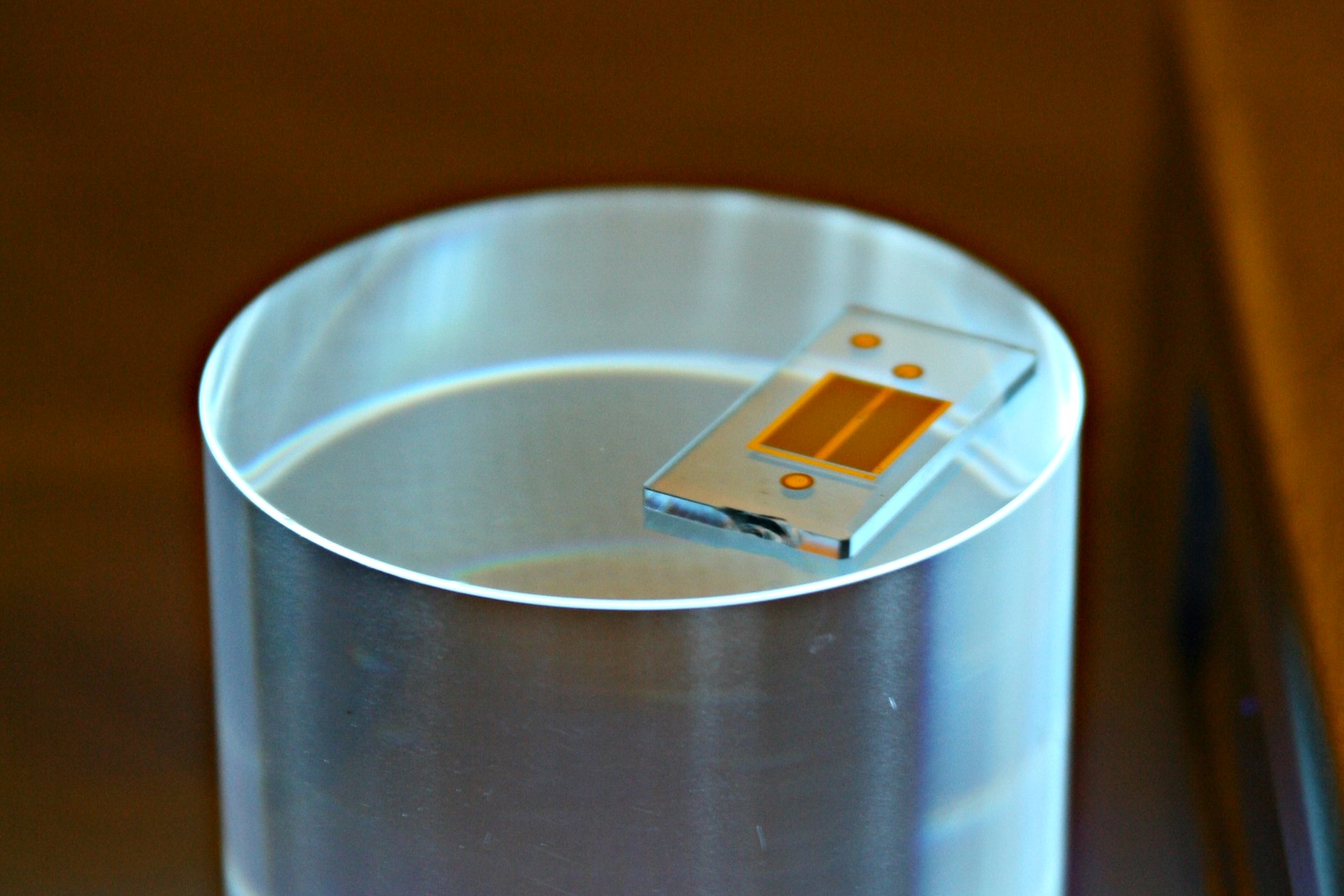}
 \caption{One of the composite phonon detectors, scintillating in UV light.}
 \label{fig:znwo}
\end{figure}

The concept of composite detectors has been previously explored, for example for GNO \cite{jean2}. For CRESST, we evaporated tungsten phonon detector thermometers onto small crystals and glued these thermometer carriers onto scintillator crystals of the same material; an example can be seen in Fig. \ref{fig:znwo}. We used two different glues; more details are described in \cite{Kiefer08}. Run 31 of the \mbox{CRESST-II} experiment included 13 modules with thermometers directly evaporated and four modules with composite detectors. Three of these were \cawo and a fourth one \znwo \cite{IrinaCryoScint}. Concerning future plans, the gluing technique might facilitate the detector mass production required for a ton-scale experiment like \mbox{EURECA} \cite{EureKraus}.\par
Results obtained with composite detectors in Run 31 are promising. The pulses of the individual crystals of a composite detector, namely the thermometer carrier and the absorber crystal, can be distinguished by their pulse shape.
\section{Experimental results}
\subsection{Setup}
The scintillating crystals used in CRESST-II have a cylindrical shape of $\unit[40]{mm}$ height and $\unit[40]{mm}$ diameter. The thermometer carriers have dimensions of $\unit[20\times10\times1]{mm^3}$. We used two different glues to connect the parts, Araldite 2011 for one detector named Hanna and EpoTek 301-2 for the other two called Maja and Rita. Both types of glue have proven to work under cryogenic conditions before. Nonetheless, after finishing Run 31, the absorber crystal of the detector Hanna was found to be destroyed by a crack reaching from the glue spot through the crystal. In spite of this fact, the energy resolution in Hanna was sufficient for dark matter search. Unfortunately, due to technical problems the light detector was not functioning so that Rita/Steven was the only composite \cawo module with a working light detector
\subsection{Calibration spectra}
\begin{figure}
  \includegraphics[width=0.9\linewidth]{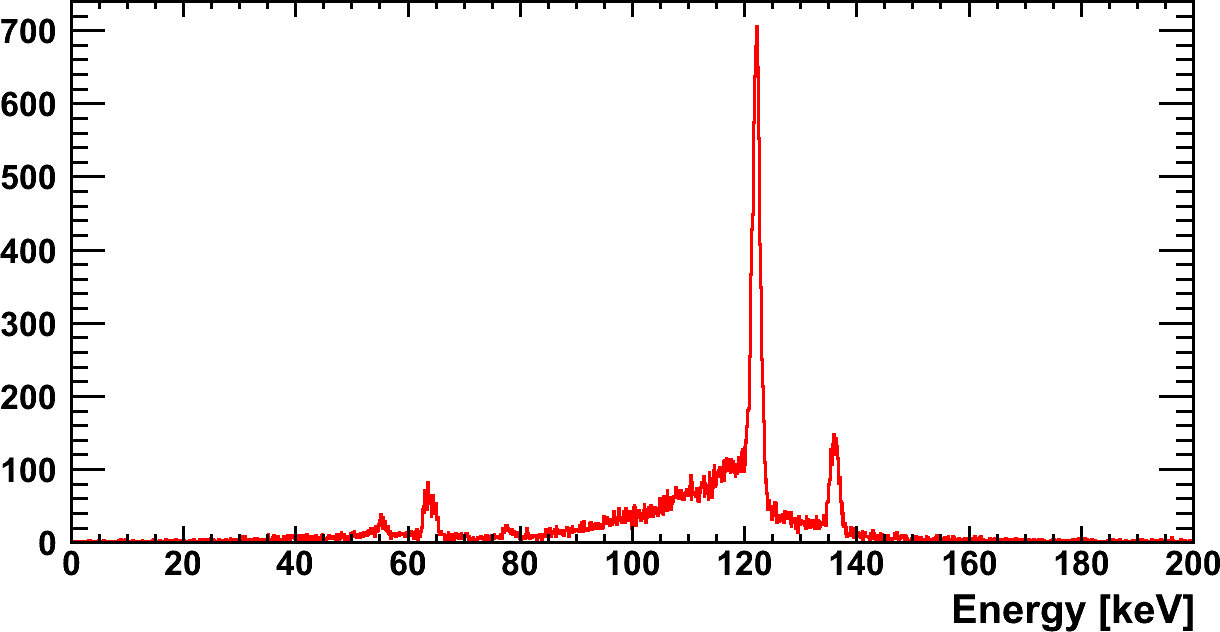}
 \caption{$^{57}$Co Calibration spectrum of the phonon detector Maja.}
 \label{fig:maja_cal}
\end{figure}
\begin{figure}
  \includegraphics[width=0.9\linewidth]{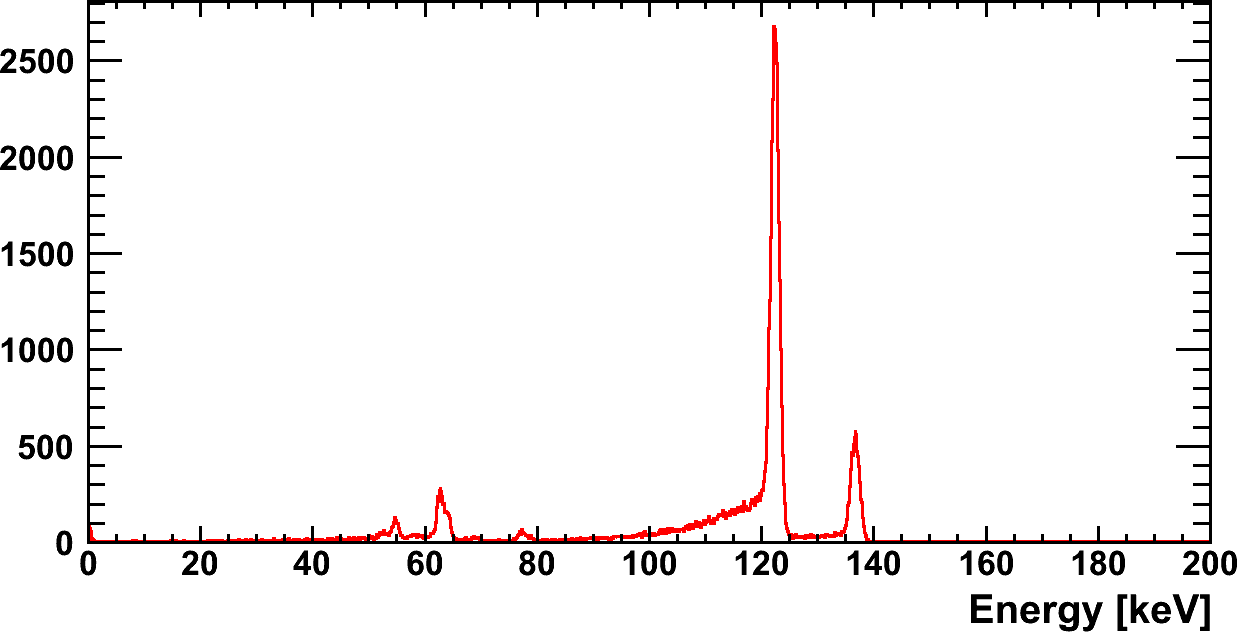}
 \caption{$^{57}$Co Calibration spectrum of the phonon detector Julia.}
 \label{fig:julia_cal}
\end{figure}
We calibrate detectors by exposing them to $\gamma$ radiation from a $^{57}$Co source. The calibration spectrum recorded with Maja can be seen in Fig. \ref{fig:maja_cal}. In addition to the lines at 122 and \unit[136]{keV}, the escape peaks of W and a Pb line can be seen as well. For comparison, Fig. \ref{fig:julia_cal} shows the calibration spectrum of Julia, a typical standard phonon detector, that we will use as a reference here. Using the FWHM of the \unit[122]{keV} line as a reference, one can compare the signal quality of the phonon detectors. \mbox{Tab. \ref{tab:FWHM}} shows the values for the three composite detectors as well as for Julia. The composite detectors perform nearly as good as Julia, with the exception of Hanna, the broken detector.
\begin{table}
\begin{tabular}{cccl}
\hline
% Name & FWHM$_{peak}~\left[\unit{keV}\right]$ & FWHM$_{baseline}~\left[\unit{keV}\right]$ & Comment\\
~ & \multicolumn{2}{c}{FWHM $\left[\unit{keV}\right]$} & ~\\
Name & $\unit[122]{keV}$-Peak & Baseline & Comment\\
\hline
Julia & 1.92 & 0.99 & standard\\
Maja & 1.98 & 0.80 & EpoTek\\
Rita & 1.98 & 0.41 & EpoTek\\
Hanna & 2.62 & 0.44 & Araldite, broken\\
\hline
\end{tabular}
\label{tab:FWHM}
\caption{Comparison of the FWHMs of the \unit[122]{keV} calibration line. Julia is a standard phonon detector while the others are composite detectors.}
\end{table}
\subsection{Light Yield}
\begin{figure}
  \includegraphics[width=0.9\linewidth]{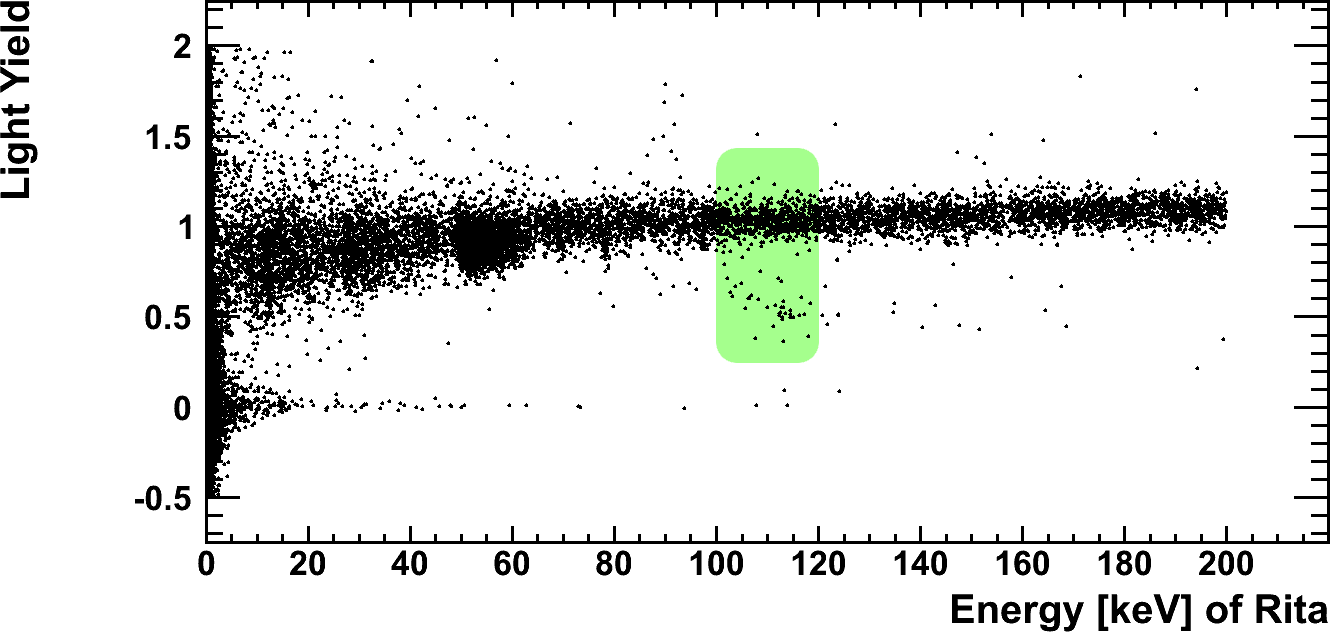}
 \caption{Light yield plot taken with the module Rita/Steven. The shaded area indicates pulses in the energy range of Pb-recoils. For a possible explanation of the events around zero light yield see \cite{LTDJens}.}
 \label{fig:rita_LY_normal}
\end{figure}
\begin{figure}
  \includegraphics[width=0.9\linewidth]{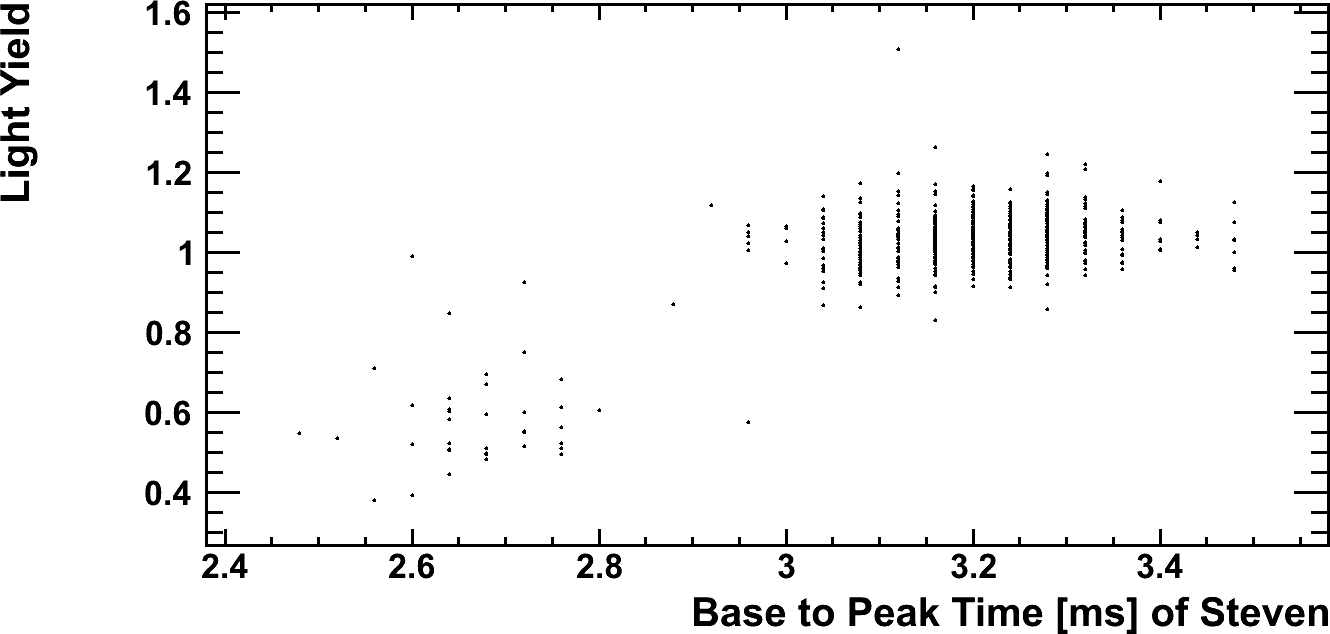}
 \caption{Time from pulse onset to peak vs. Light Yield of the light pulses around \unit[100]{keV} in Rita/Steven. The different times indicate a fast scintillator as the origin of the events with the lower light yield.}
 \label{fig:rita_PeakPos}
\end{figure}
In order to find out about the nature of a particle event, the data of the light detector and the phonon detector have to be combined. This results in a plot where the particle energy, recorded in the phonon detector, is shown on the abscissa. The pulse height of the light signal divided by the particle energy is shown on the ordinate. This so-called Light Yield is normalized to 1 for $\gamma$ particles at \unit[122]{keV}. Fig. \ref{fig:rita_LY_normal} shows such a plot, taken with the phonon detector Rita and its corresponding light detector Steven. In the shaded region between 100 and \unit[120] keV, there is a population of particle events below the $\gamma$ band.
These events originate from $^{210}$Po $\alpha$-decays where the $\alpha$-particle is absorbed in the surrounding scintillating reflector foil and the recoiling nucleon hits the \cawo crystal \cite{lang2009a}. The light signals show a faster rise time consistent with the fast scintillation time of the plastic reflector foil (see Fig. \ref{fig:rita_PeakPos}). As the same foil is used in every detector module, its absolute light yield is the same for every module. Hence it is possible to use the relation of the Light Yield of the $\gamma$ band and of this spot as a mean for comparing the light output of different scintillator crystals without having installed a dedicated light detector calibration source in each module. By using this method however, there was no significant difference between the detectors Rita and Julia. Analysis of data gathered in Run 32, where the energy scale of the detector was calibrated by shining \unit[6]{keV} X-rays onto the light detector, resulted in a light output of \unit[2.36]{\%} for Rita which is an extraordinarily high value, compared to standard detectors, where the value is around \unit[1.3]{\%}. The reason for this discrepancy remains to be clarified.
\subsection{Pulse shapes}
\begin{figure}
  \includegraphics[width=0.9\linewidth]{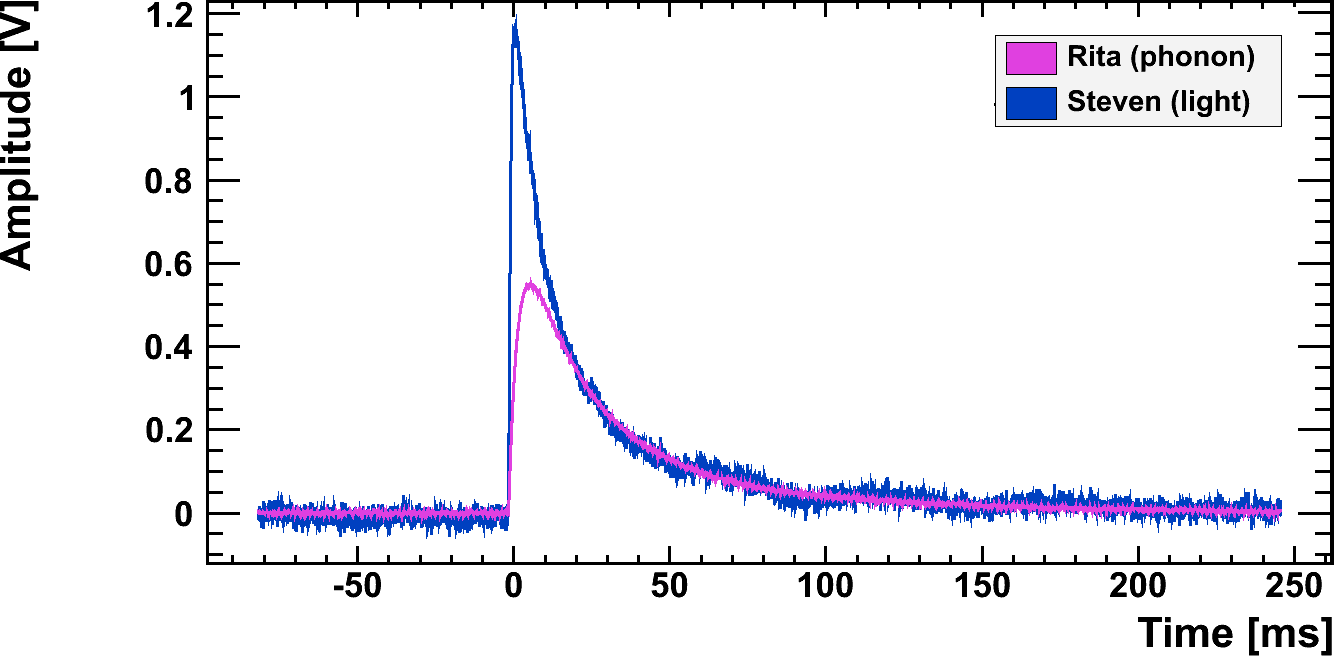}
 \caption{$\gamma$ event taken with the module Rita/Steven. The energy was absorbed in the absorber crystal.}
 \label{fig:rita_part_normal}
\end{figure}
\begin{figure}
  \includegraphics[width=0.9\linewidth]{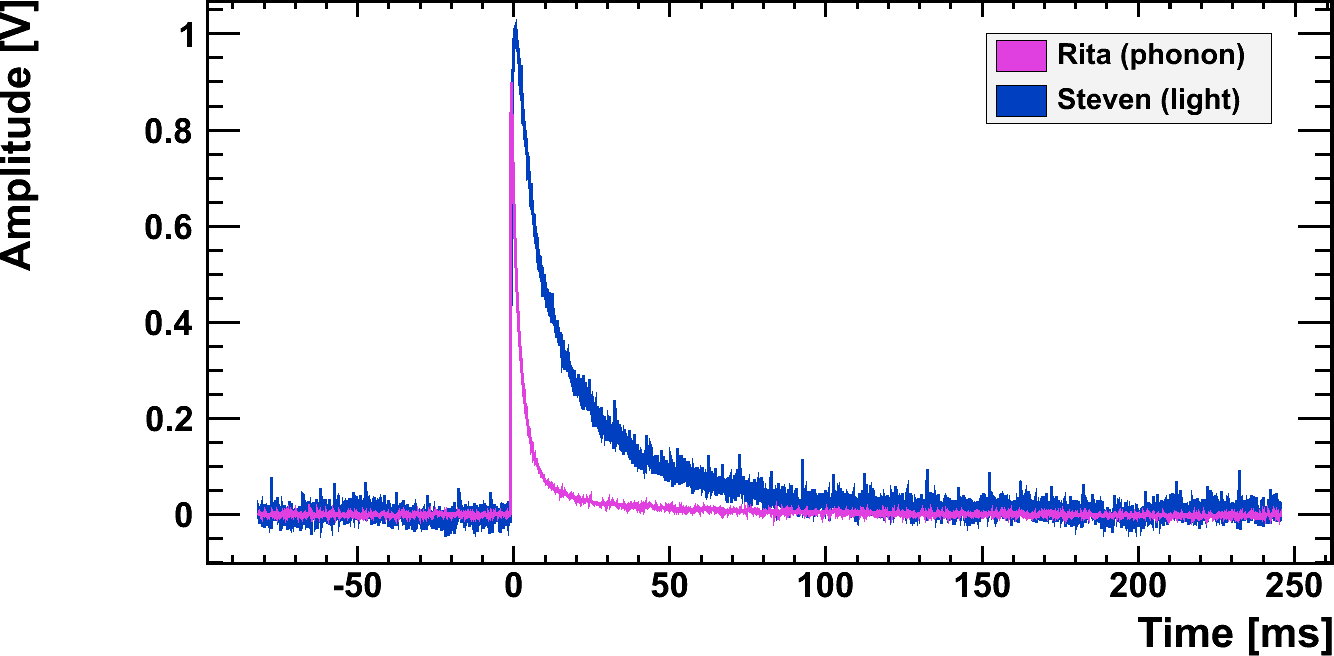}
 \caption{$\gamma$ event taken with the module Rita/Steven. The energy was absorbed in the thermometer carrier crystal.}
 \label{fig:rita_part_spike}
\end{figure}
Fig. \ref{fig:rita_part_normal} shows a $\gamma$ event recorded by the module Rita/Steven. The pulse can be described as a sum of three exponentials \cite{proebst}. When comparing this pulse with the one shown in Fig. \ref{fig:rita_part_spike}, one sees that the rise and decay times of the latter are much shorter. The first pulse was caused by an energy deposition in the absorber crystal while in the case of the second pulse, the deposition occurred directly in the thermometer carrier (cf. \cite{Kiefer08,Roth2008}).
\section{Further research}
\begin{figure}
  \includegraphics[width=0.45\linewidth]{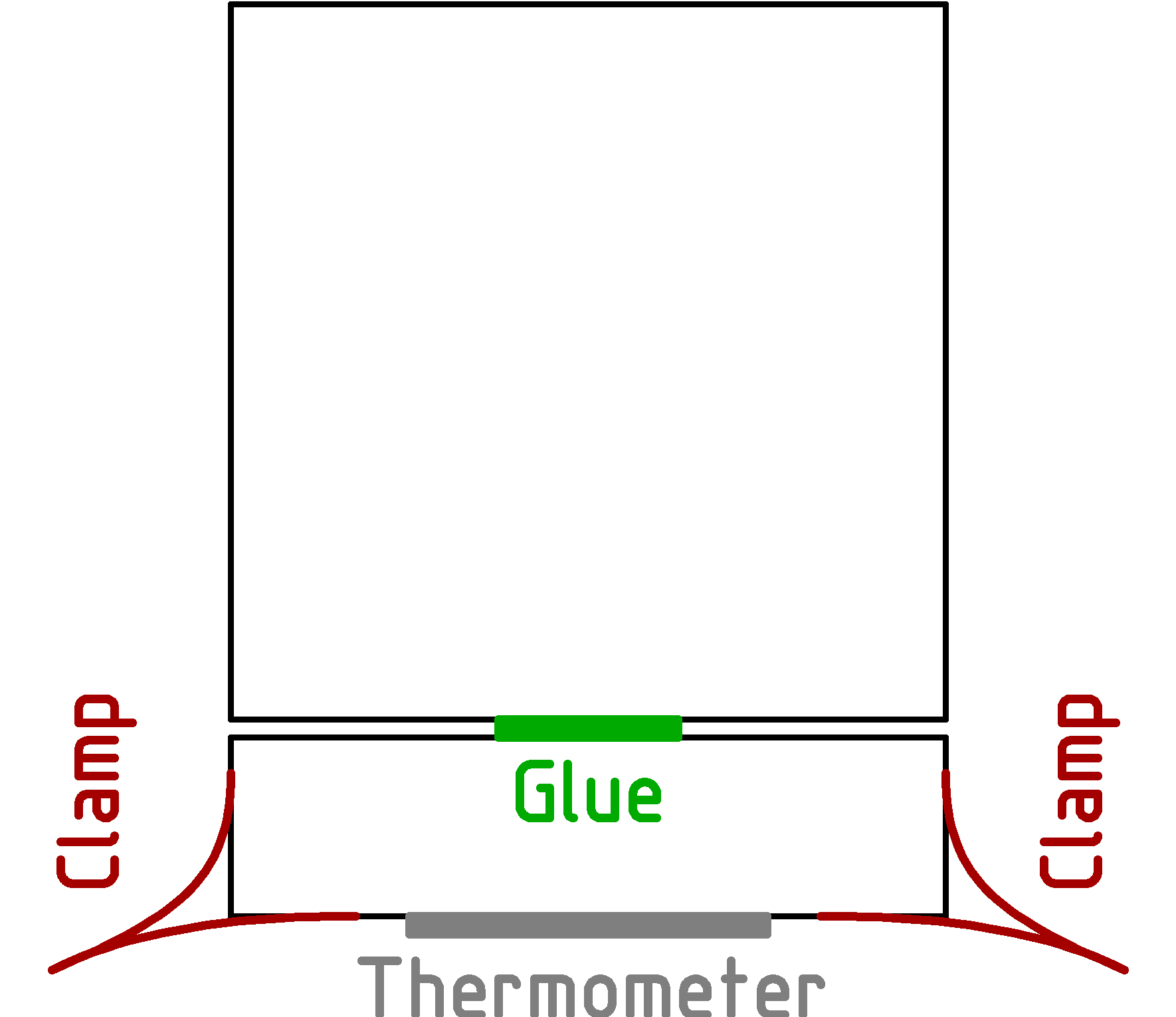}
 \caption{Schematic concept of the clamp-event discrimination detector}
 \label{fig:clampdet}
\end{figure}
In spring of 2009, CRESST started a new run. \mbox{Run 32} features 13 standard \cawo detectors, two composite \cawo and two composite \znwo detectors. Additionally, there is a detector which is designed to discriminate events caused by the holding clamps from the events created in the absorber crystal. We have used EpoTek for all composite detectors.\par
This detector consists of a thermometer film on a \cawo disk of \unit[40]{mm} diameter and \unit[10]{mm} height which is glued to a cylindrical absorber crystal of $\unit[40]{mm}\times\unit[40]{mm}$ (see Fig. \ref{fig:clampdet}). In contrast to a standard phonon detector, this one is only held by clamps at the smaller disk. The clamp-event discrimination detector takes advantage of the different pulse shapes in order to separate events occurring in the clamps \cite{LTDJens} from events occurring in the absorber: The events coming from the clamps should have a pulse shape similar to the events originating from the small part and thus should be distinguishable from events originating in the big absorber crystal.
\section{Conclusion}
The composite detectors included in Run 31 have generated valuable data. Measurements indicate that the composite detectors might have an advantage in light output, compared to conventional detectors. There is a possibility to use other, more delicate materials such as \znwo now, that they can be equipped with thermometers which was not possible before. Additionally, the different-sized crystals create pulses with differnt shapes, providing information about the origin of a recorded event. We will take advantage of this in Run 32, where a special detector is included in order to investigate in pulses created by the holding clamps.
\section{Acknowledgements}
This work was partially supported by funds of the DFG Transregio 27 ``Neutrinos and Beyond'', the Munich cluster of Excellence ``Origin and Structure of the Universe'' and the Maier-Leibnitz-Laboratorium. Support from the Science and Technology Facilities Council (STFC), UK is acknowledged.
\bibliographystyle{aipproc}
\bibliography{LTD2009-kiefer}
\end{document}